\documentclass[conference]{IEEEtran}

\begin{document}
%
\title{ Improving abcdSAT by At-Least-One Recently Used Clause Management Strategy}

\author{\IEEEauthorblockN{Jingchao Chen}
\IEEEauthorblockA{School of Informatics, Donghua University\\
2999 North Renmin Road, Songjiang District, Shanghai 201620, P. R. China\\
chen-jc@dhu.edu.cn}}

\maketitle

\begin{abstract}
 We improve further the 2015 version of abcdSAT by various heuristics such as
 at-least-one recently used strategy, learnt clause database approximation reduction etc.
 Based on the requirement of different tracks at the SAT Competition 2016, we develop three
 versions of abcdSAT: \emph{drup}, \emph{inc} and \emph{lim}, which participate in
the competition of  main (agile), incremental library and no-limit track, respectively.

\end{abstract}

\section{Introduction}

    The abcdSAT solver submitted to the SAT Competition 2016 is the improved version of abcdSAT 2015 \cite{abcdSAT15},
    which are built on the top of Glucose 2.3 \cite{glucose2.3,glueLBD}.
Here we provide three versions of abcdSAT: \emph{drup}, \emph{inc} and \emph{lim}, which are submitted to
main (agile) track, incremental library track and no-limit track, respectively. The main techniques use by the three
versions include at-least-one recently used strategy, learnt clause database approximation reduction,
recursive splitting solving, decision variable selection based on blocked clause decomposition \cite{MixBCD,SATBCD},
bit-encoding phase selection \cite{BitPhase}, simplification such as
lifting, probing, distillation, elimination, hyper binary resolution etc. Of course, all the simplification techniques
used here are the existing techniques, so we will omit the description on them.

\section{At-Least-One Recently Used Strategy}

In the search process of CDCL (Conflict Driven, Clause Learning) solvers, the learnt clause database is required to be maintained.
Based on our experimental observation, the clause database maintenance is actually similar to cache replacement in CPU cache management or
page replacement in a computer operating system. There are many
cache (page) replacement algorithms. For example, Least Recently Used (LRU), Most Recently Used (MRU),
Pseudo-LRU (PLRU), Least-Frequently Used (LFU), Second Chance FIFO, Random Replacement (RR), Not Recently Used (NRU) \cite{LRU} etc.
Our At-Least-One Recently Used (ALORU) algorithm is similar to NRU page replacement algorithm, but
different from the clause freezing mechanism proposed by Audemard et al \cite{freeze}.
ALORU favours keeping learnt clauses in database that have been recently used at least one time. If a learnt clause
has not so far involved in any conflict analysis since it was generated, it will be discarded first. Implementing ALORU
is very simple. When a conflict clause (called also learnt clause) is generated, its LBD (Literal Block Distance, for its definition, see \cite{glueLBD})
is usually set to the number of different decision
levels involved in it. However, ALORU sets the initial LBD of a conflict clause to $+\infty$, not actual current LBD.
In details, in the {\bf search} procedure, ALORU replaces ``{\bf setLBD(nblevels)}" with ``{\bf setLBD(0x3fffffff)}".
Since any LBD in real instances that can be solved never exceeds 0x3fffffff, we denote $+\infty$ with 0x3fffffff. If a learnt clause involves in a conflict analysis,
the procedure for conflict analysis sets its LBD to the actual value.

\section{Learnt Clause Database Approximation Reduction}

  The target of learnt clause database reduction is two fold: remove useless clauses and avoid the expansion of database.
  Almost all the existing reduction algorithms in CDCL solvers are to sort learnt clauses according to the score (e.g. LBD) of clauses, then remove
  a given number of clauses in the sorted order. This can be viewed as exact reduction. Our approximation reduction is different
  from the exact reduction. It has no sorting, and replaces sorting with selection. In details, our approximation reduction finds
firstly the clause with the $k$-th smallest (or largest) score, where $k$ is the number of clauses to be removed. Secondly, it removes $k$ clauses with the score less than or equal
to the $k$-th smallest score. Notice, the clauses with the score equal
to that of the $k$-th clause are not often unique. And the clauses with the score less than to the $k$-th smallest score are not necessarily removed.
Therefore, the parameter $k$ is an approximation value or estimate, not exact. Due to this reason, we call reduction implemented by finding the $k$-th item
approximation reduction. Here we choose QUICKSELECT or Hoare's FIND algorithm \cite {quickselect} to find the $k$-th item.

   If all database reductions are done in this approximation way, solving is not the most efficient. Therefore, we apply the approximation reduction when
   the number of conflicts is larger than 300000 for special CNF instances. In the other cases, we apply still the exact approximation.

\section{Dynamic Core and Local Learnt Clause Management}

Like SWDiA5BY \cite{coreLearnt}, glue\_alt classifies also learnt clauses
into two categories: core and local. However, the classification of SWDiA5BY is static,
while our classification is dynamic. In SWDiA5BY, the maximum LBD of core learnt clauses is
fixed to a constant 5. However, in abcdSAT, the maximum LBD of core learnt clauses is
not fixed. AbcdSAT divides the whole search process two stages. When the number of conflicts is less than $2 \times 10^6$,
it is considered as the first stage. Otherwise, it is considered as the second stage.
In the first  stage, the maximum LBD of core learnt clauses is limited to 2. At this stage, core learnt clauses
are kept indefinitely, unless eliminated when they are satisfied. In the second stage, the maximum
LBD of core learnt clauses is limited to 5. This stage does not ensure that core learnt clauses
are kept indefinitely. When  local learnt clause database is reduced, we move 5000 core learnt clauses with
LBD larger than or equal to 3 to local learnt clause database.

   Whether the first or second stage, the number of local learnt clauses
   is maintained roughly between 9000 and 24000. That is, once the number of local learnt clauses reaches
   a upper bound, say 18000, abcdSAT will halve the number of the clauses. And the clauses with the smallest
   activity scores are removed first. The computation of clause activity scores here is consistent
   with MiniSat.

\section{Recursive Splitting Solving}

   Any CNF formula $\cal F$ can be split into two subproblems ${\cal F} \cup \{x\}$
 and ${\cal F} \cup \{\neg x\}$, where $x$ is a variable in $\cal F$.
 We can obtain the solution the original problem by solving each subproblem.
 Solving subproblem in the same way results in a recursive solving algorithm.
 In general, we limit recursive depth to 10.  Here is the pseudo-code of this recursive solving
framework.

\begin{small}
\begin{flushleft}
{\bf Algorithm } SplitSolve($\cal F$, $level$)\\
\hskip 4mm   {\bf if} $level \geq 10$ {\bf then return} abcdSAT($\cal F$, $2\times 10^6$)\\
\hskip 4mm   $\langle ret, \cal F' \rangle$ $\leftarrow$ abcdSAT($\cal F$, 500)\\
\hskip 4mm   {\bf if} $ret=$ SAT or UNSAT {\bf then return} $ret$\\
\hskip 4mm  $x$ $\leftarrow$ GetBranchVariable($\cal F'$)\\
\hskip 4mm  SplitSolve(${\cal F'} \cup \{x\}$, $level+1$)\\
\hskip 4mm  SplitSolve(${\cal F'} \cup \{\neg x\}$, $level+1$)\\
\end{flushleft}
\end{small}

The 2nd parameter of abcdSAT in the above algorithm denotes the limit of the number of conflicts.
abcdSAT($\cal F$, 500) means that it searches a solution of $\cal F$ until the number of conflicts reaches 500.
Procedure GetBranchVariable selects a branch variable according to the rule given in \cite {hybridsolve}.

 This solving framework is suitable for small formulas.

\section{abcdSAT \emph{drug}}

Because each solver participating in the main track is required to provide a DRUP proof in UNSAT case,
we add a DRUP patch in the original abcdSAT. In addition to this patch, abcdSAT \emph{drug} adds
learnt clause database approximation reduction, at-least-one recently used strategy, but excludes XOR and
cardinality constraint simplification. In particular, XOR simplification is difficult to provide a DRUP proof.
The splitting and merging technique used in the original abcdSAT cannot provide a DRUP proof. So abcdSAT \emph{drug} simplifies
it into recursive splitting solving technique given in previous section.

\section{abcdSAT \emph{inc}}

 The solver submitted to the incremental library track is called abcdSAT \emph{inc}. This version has no DRUP patch.
The main difference between abcdSAT \emph{inc} and abcdSAT \emph{drug} is that abcdSAT \emph{inc} adopts
dynamic core and local learnt clause management policy, while abcdSAT \emph{drug} adopts Glucose-style learnt clause management policy.

\section{abcdSAT \emph{lim}}

This is the version submitted to the no-limit track. AbcdSAT \emph{lim} not only includes various techniques given
above, but also XOR and cardinality constraint simplification. With respect to learnt clause management, what abcdSAT \emph{lim} adopts
is Glucose-style learnt clause management policy. For a few special instances, abcdSAT \emph{lim} switches to Lingeling 587f \cite{Lingeling:587f}
to solve them. When the average LBD score of an instance to be solved is small, say less than 16,  this version uses splitting and
merging (reconstructing) strategy described in \cite{splitMerge}, rather than recursive splitting solving strategy mentioned above.


\begin{thebibliography}{1}

\bibitem{abcdSAT15}
J.C.~Chen: MiniSAT\_BCD and abcdSAT: solvers based on blocked clause decomposition,
in \emph{Proceedings of the SAT Competition 2015}

\bibitem{MixBCD}
J.C.~Chen: Fast Blocked Clause Decomposition with High Quality, 2015, http://arxiv.org/abs/1507.00459

\bibitem{SATBCD}
Chen, J.C.: Improving SAT Solvers via Blocked Clause Decomposition, 2016,
http://arxiv.org/abs/1604.00536

\bibitem{BitPhase}
J.C.~Chen:A bit-encoding phase selection strategy for satisfiability
  solvers,in \emph{Proceedings of Theory and Applications of Models of
  Computation ({TAMC}'14)}, ser. LNCS 8402, 2014, pp. 158--167.

\bibitem{splitMerge}

J.C.~Chen: Glue\_lgl\_split and GlueSplit\_clasp with a Split and Merging Strategy,
in \emph{Proceedings of the SAT Competition 2014}, pp. 37--39.

\bibitem{glucose2.3}
G.~Audemard and L.~Simon, Glucose 2.3 in the sat 2013 competition,in
  \emph{Proceedings of the SAT Competition 2013}, pp. 40--41.

\bibitem{glueLBD}
G.~Audemard, L.~Simon:Predicting learnt clauses quality in modern sat
  solvers, in \emph{proceedings of IJCAI}, 2009, pp. 399--404.

\bibitem{freeze}
G.~Audemard, J.M.~Lagniez, B.~Mazure, L.~Sa\"{i}s:On freezing and reactivating learnt clauses,
in \emph{Proceedings of SAT 2011}, ser. LNCS, vol. 6695, pp. 188--200.

\bibitem{LRU}
Amit S. Chavan, Kartik R. Nayak, Keval D. Vora, Manish D. Purohit, Pramila M. Chawan: A comparison of page
replacement algorithms, IACSIT, vol.3, no.2, April 2011.

\bibitem{quickselect}
C.~Hoare, Algorithm 63 (PARTITION), Algorithm 64 (QUICKSORT) and Algorithm 65 (FIND). Communications of
the ACM 1961,4(7), pp. 321--322.

\bibitem{coreLearnt}
C., Oh: MiniSat HACK 999ED, MiniSat HACK 1430ED, and SWDiA5BY, in \emph{Proceedings of the SAT Competition 2014}, pp. 46--47.

\bibitem{hybridsolve}
J.C.~Chen: Building a hybrid sat solver via conflict-driven, look-ahead and xor
  reasoning techniques,in \emph{Proceedings of SAT 2009}, ser. LNCS, vol.
  5584, pp. 298--311.

\bibitem{Lingeling:587f}
A.~Biere: Lingeling, plingeling and treengeling. [Online]. Available:
  http://fmv.jku.at/lingeling/

\end{thebibliography}
\end{document}